\documentclass[letterpaper]{article} %
\usepackage{aaai24}  %
\usepackage{amsmath}
\usepackage{amsthm}
\usepackage{booktabs}
\usepackage{multirow}
\usepackage{xcolor}
\usepackage{microtype}
\usepackage{tikz}
\usepackage{tikzscale}
\usepackage{pgfplots}
\usepackage{tabularray}

\definecolor{myred}{RGB}{241, 156, 153}
\definecolor{mygreen}{RGB}{185, 224, 165}

\newcommand{\system}[0]{\texttt{FLAME}}
\usepackage{subcaption}

\usepackage{cleveref} %
\crefname{question}{RQ}{RQs}

\newcommand{\aaai}[0]{Forum}

\newtheorem{example}{Example}

\usepackage{times}  %
\usepackage{helvet}  %
\usepackage{courier}  %
\usepackage[hyphens]{url}  %
\usepackage{graphicx} %
\usepackage{natbib}  %
\usepackage{caption} %
\usepackage{algorithm}
\usepackage{algorithmic}

\usepackage{newfloat}
\usepackage{listings}
\DeclareCaptionStyle{ruled}{labelfont=normalfont,labelsep=colon,strut=off} %
\floatstyle{ruled}
\newfloat{listing}{tb}{lst}{}
\floatname{listing}{Listing}
\title{FLAME: A Small Language Model for Spreadsheet Formulas}
\author {
    Harshit Joshi\textsuperscript{\rm 1}\thanks{Work done while at Microsoft},
    Abishai Ebenezer\footnotemark[1],
    Jos\'e Cambronero Sanchez\textsuperscript{\rm 2}\thanks{Listed in alphabetical order},
    Sumit Gulwani\textsuperscript{\rm 2}\footnotemark[2],\\
    Aditya Kanade\textsuperscript{\rm 3}\footnotemark[2],
    Vu Le\textsuperscript{\rm 2}\footnotemark[2],
    Ivan Radi\v{c}ek\footnotemark[2]\footnotemark[1],
    Gust Verbruggen\textsuperscript{\rm 2}\footnotemark[2]
}
\affiliations {
    \textsuperscript{\rm 1} Stanford University,
    \textsuperscript{\rm 2} Microsoft, USA,
    \textsuperscript{\rm 3} Microsoft Research, India\\
    harshitj@cs.stanford.edu, \{abishai.ebenezer.m, ivan.radicek\}@gmail.com, \{jcambronero, sumitg, kanadeaditya, levu, gverbruggen\}@microsoft.com
}

\begin{document}

\maketitle

\begin{abstract}
Spreadsheets are a vital tool for end-user data management.
Using large language models for formula authoring assistance in these environments can be difficult, as these models are expensive to train and challenging to deploy due to their size (up to billions of parameters). 
We present \system{}, a transformer-based model trained exclusively on Excel formulas that leverages domain insights to achieve competitive performance while being substantially smaller (60M parameters) and training on two orders of magnitude less data.
We curate a training dataset using sketch deduplication, introduce an Excel-specific formula tokenizer, and use domain-specific versions of masked span prediction and noisy auto-encoding as pre-training objectives.
We evaluate \system{} on formula repair, formula completion, and
similarity-based formula
retrieval.
\system{} can outperform much larger models, such as the Davinci (175B) and Cushman (12B) variants of Codex and CodeT5 (220M), in 10 of 14 evaluation settings for the repair and completion tasks.
For formula retrieval, \system{} outperforms CodeT5, CodeBERT, and GraphCodeBERT.
\end{abstract}

\begin{figure}[!t]
    \centering
    \includegraphics[width=\linewidth]{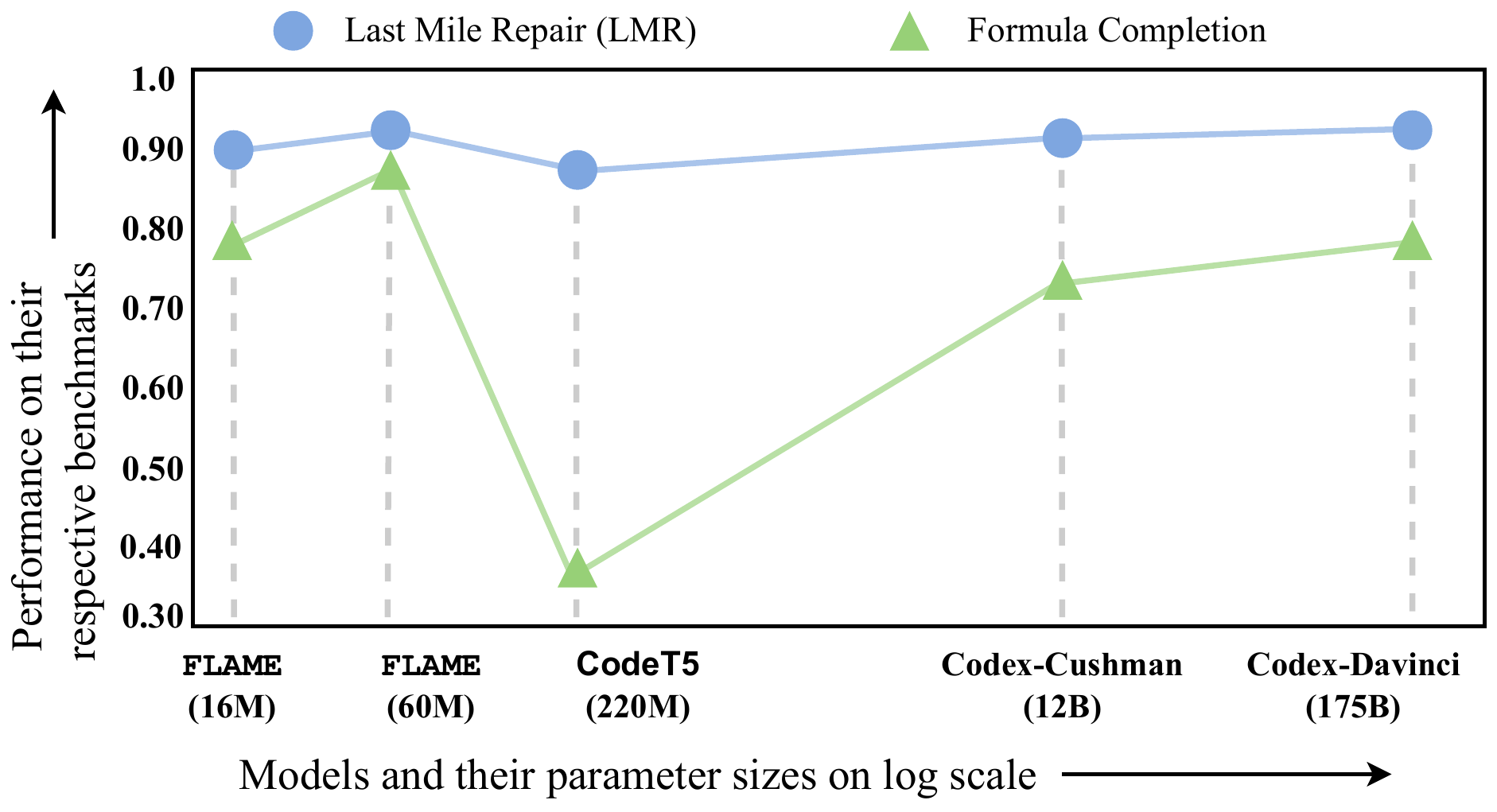}
    \caption{Model performance (top-5 candidate cutoff) for last-mile repair
    and completion on the benchmark introduced by ~\cite{ring} and our new benchmark for completion. Note that Codex-Davinci results are few-shot, and completion is zero-shot for all systems except CodeT5.
    Completion results are the fraction of benchmarks successfully completed (based on sketch match metric) given a 90\% prefix.
    }
    \label{fig:intro-summary-plot}
\end{figure}
\section{Introduction}
Spreadsheets remain an important data management and processing tool for end-users~\cite{rahman2020benchmarking} and estimates of the number of spreadsheet users are in the billions~\cite{morgan-stanley-call}. 
However, despite a large user base, spreadsheet environments still do not  have access to nearly the same range of productivity tools available for general programming environments.
The latter typically have code completion, refactoring, linting, and a wide range of extensions for additional functionality, like generating tests, inserting code snippets, and summarizing code.
Many of these recent advancements in programming assistance tools
are driven by large language models (LLMs) trained on code \cite{codex,polycoder,incoder,codegen}, and include features
for code completion \cite{copilot}, repair \cite{ring}, and automated reviews \cite{li2022automating}.

To capture the complexity and variety of code and comments in different languages, these models need billions of parameters (Codex has 12B and 175B variants), are trained for long periods of time
on millions of programs (Incoder was trained on 159GB of code for 24 days), and result in expensive inference due to hardware requirements.
For tasks constrained to a spreadsheet formula language without natural language interaction, such as repairing a broken spreadsheet formula, a substantial amount of such models' expressive power is left unused.
This raises the question: \textit{can we substantially reduce the model size by focusing on  spreadsheets and exclusively on the formula-language used to carry out computations in such an environment?}

In this paper, we present \system{}, a \texttt{F}ormula \texttt{LA}nguage \texttt{M}odel for \texttt{E}xcel trained exclusively on Excel formulas.
\system{} is based on T5-small~\cite{t5} and has only 60 million parameters, yet it can compete with much larger models (up to 175B parameters) on last-mile formula repair and formula completion.
Additionally, we compare \system{} to CodeT5, CodeBERT, and GraphCodeBERT on retrieval of similar formulas.

\begin{figure}[!t]
    \centering
    \includegraphics[width=\linewidth]{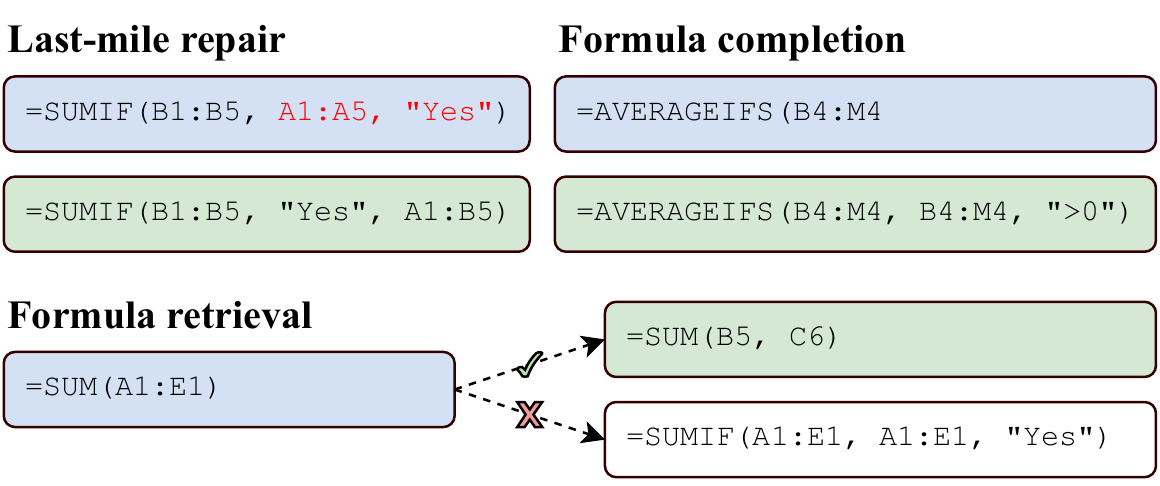}
    \caption{We experiment on three downstream tasks: last-mile repair, formula completion, and retrieving similar formulas. Blue indicates the input and green indicates the expected output. Red text denotes the buggy part of the formula in the repair task, where the user has swapped the correct order of arguments resulting in a type error. \system{} succeeds in all of these cases.
    }
    \label{fig:downstream_tasks}
\end{figure}

Figure~\ref{fig:intro-summary-plot} shows a summary of results as a function of model on a public dataset, where \system{} can outperform or compete in both 
formula completion and repair tasks.
Figure~\ref{fig:downstream_tasks} shows some examples, solved by \system{}, for these tasks.

\begin{figure*}[!t]
    \centering
    \includegraphics[width=\linewidth]{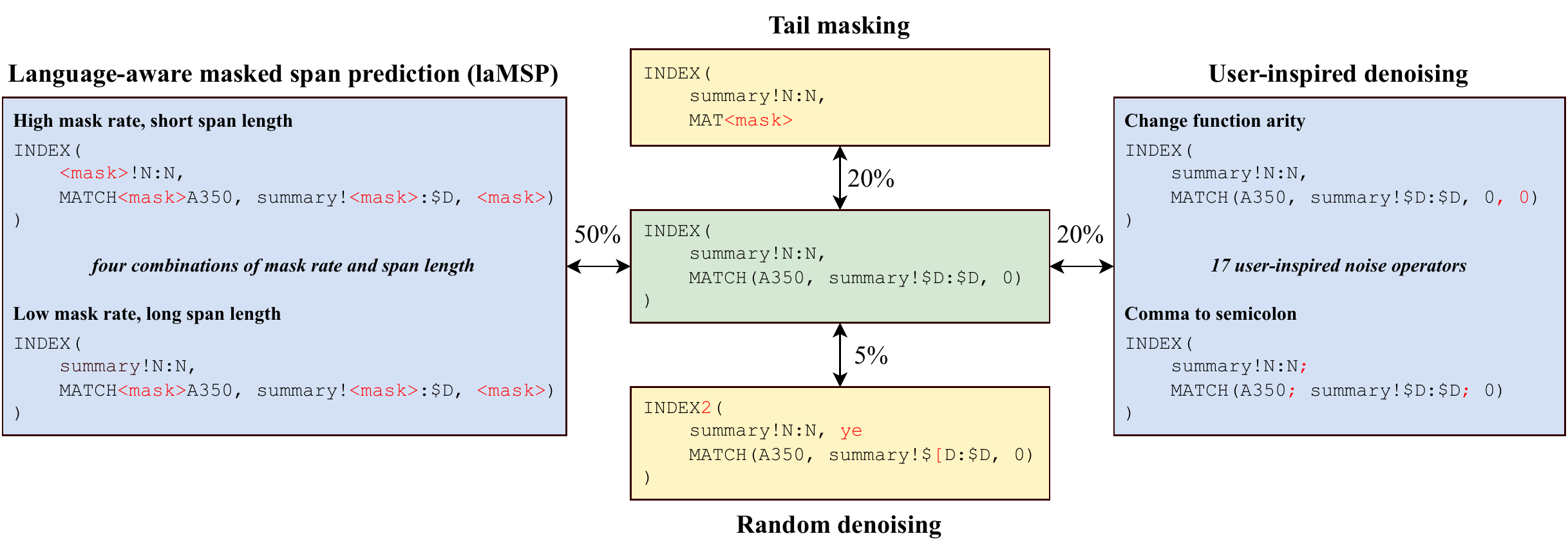}
    \caption{
    Four pre-training objectives used by \system{}.
    For each batch, we randomly (with weighted probability) sample one of the four objectives.
    Generic objectives (tail masking and random noise) are shown in yellow, formula-specific variants (language-aware span masking and user-inspired noise) are shown in blue.
    For all pre-training objectives, \system{} needs to generate the complete original formula (green).}
    \label{fig:pretraining-objectives}
\end{figure*}

There are three main challenges for training a model on Excel formulas: obtaining diverse training data, tokenizing formulas' unique structure, and designing pre-training objectives that teach the model about this structure. 

To create a small but varied pre-training corpus, we reduced a corpus of 927M formulas down to 6.1M by comparing formulas based on syntax, removing near-duplicate formulas, which only vary in terms of cell inputs (a common occurrence due to the grid nature of the environment).
To account for formula structure, we combine formula-specific tokenization with byte-pair encoding (BPE) to train a tokenizer.
To incorporate formula structure into pre-training, we introduce two new domain-specific pre-training objectives: language-aware masked span prediction and user-inspired denoising, which complement two generic objectives (tail-masking and denoising auto-encoding).

We evaluate \system{} on formula completion and repair, showing that  \system{} can significantly improve over general code models and can compete with much larger models.
Specifically, \system{} outperforms other models in 10 of 14 settings in our repair and completion evaluation. In addition, we also show that \system{} embeddings can be used to retrieve similar formulas more efficiently and effectively than CodeT5, CodeBERT, and GraphCodeBERT.

We make the following contributions:
\begin{itemize}
    \item We present \system{}, the first language model designed exclusively for Excel formulas.
    To this end, we introduce domain-specific dataset curation, tokenization, and pre-training objectives.
    \item We extensively evaluate \system{} and other larger language models on three formula assistance tasks: last-mile repair, formula completion, and formula retrieval.
    \item We analyze the impact of deduplication, tokenization, and training objectives
     on \system{}.
\end{itemize}
\section{Related Work}

Multiple language model architectures have been 
successfully adapted to code to produce popular
models such as CodeBERT~\cite{codebert},
CodeT5~\cite{codet5}, Codex~\cite{codex},
PolyCoder~\cite{polycoder} and others.
These models are trained on multiple programming languages and use 
language-agnostic
pre-training objectives.
In contrast, \system{} focuses on a single domain and uses domain-specific objectives.
BART~\cite{bart} and UL2~\cite{ul2} introduced similar objectives in the domain of natural language.

To train \system{}, we perform syntax-aware deduplication within workbooks to avoid repeating similar formula structures typical in spreadsheet environments. 
\citet{allamanis2019adverse} highlighted the related problem of duplication in traditional sources used for training machine learning models for code.

We evaluate \system{}'s ability to perform
last-mile formula repair~\cite{lamirage},
completion, and similar formula retrieval.
Prior work in the repair domain includes
DeepFix~\cite{deepfix}, BIFI~\cite{bifi}, Dr.Repair~\cite{drrepair},
TFix~\cite{tfix}, and RING~\cite{ring}, which use deep learning to perform syntax, compilation, or diagnostics repair in general-purpose programming languages.
Popular general autocompletion systems include
GitHub Copilot in VS Code~\cite{copilot}. There is a long history of language models
being used for retrieval tasks~\cite{tao2006language,zhao2009proximity, song1999general, gao2004dependence}. This work is complementary as \system{} is exclusively designed for spreadsheet environments and is only trained on formulas (and not the associated data context).

In the spreadsheet domain, SpreadsheetCoder \cite{spreadsheetcoder} predicts simple formulas from the data context.
HermEs \cite{he2023hermes} leverages a table encoder (TuTA) to 
generate a formula using
hierarchical decoding.
Previous work has also shown that pre-training over table and code data can improve table task performance~\cite{dong2022table, singh2023format5}. This body of work trains on spreadsheet table contents for tasks such as in-context formula prediction or general
spreadsheet intelligence.
In contrast, \system{} exclusively uses formulas and pre-trains a general formula model that we fine-tune for
tasks like repair.
\section{Approach}
\label{sec:approach}

We now describe the \system{} architecture and pre-training process (training data, tokenization and objectives).

\subsection{Architecture}
\label{approach:architecture}

Encoder models like CodeBERT \cite{codebert} show remarkable code understanding capabilities.
Decoder models like CodeGen \cite{codegen} and Codex~\cite{codex} perform well on code generation.
Encoder-decoder models seek to blend these strengths.
To facilitate both formula understanding and generation, \system{} uses the T5~\cite{t5} encoder-decoder architecture.

\subsection{Training Data}
\label{approach:trainingdata}
We start with a dataset of 927M formulas drawn from a corpus of 1.8M public Excel workbooks collected from the web.\footnote{Publicly available sources of spreadsheets that can be used as alternatives include Enron~\cite{enron}, EUSES~\cite{euses}, and FUSE~\cite{barik2015fuse}.}
Each workbook contains one or more worksheets, and each worksheet contains zero or more formulas.
Formulas are often repeated with only cell reference changes.

We compute formula sketches to preserve a single instance of each unique formula per workbook.
In a formula sketch, numeric constants, string constants and cell references are replaced by their token type. 
For example, the sketch of \texttt{=SUM(A1:A10)} is \texttt{=SUM(cell:cell)}.
After applying sketch deduplication, we are left with 6.1M formulas.
Note that applying this globally to the corpus, rather than per workbook, results in only 591K formulas.
We found this globally de-duplicated corpus to be insufficient for training as it skews the distribution of formulas (see Evaluation).

\subsection{Tokenizing Formulas}
\label{approach:tokenization}

A popular method for tokenization of code for neural models is byte pair encoding (BPE) \cite{bpe}, which iteratively joins consecutive tokens that appear together most frequently until a target vocabulary size is reached.
BPE is appealing for the formula domain as it can mitigate the variety in natural language content, such as string literals, named tables, column names, and sheet names. 
However, applying this procedure without modification can have adverse effects on formulas.
For example, in our corpus, \texttt{SUM} and \texttt{(} are combined to get \texttt{SUM(}, which can reduce expressiveness and hurt performance for tasks like repair.

Our tokenizer considers punctuation, whitespace, built-in function names, and digits as individual tokens~\cite{chowdhery2022palm}. 
These are often the main lexical marker between separate Excel tokens. %
We then apply BPE to the remaining parts of formulas like string constants, table names, column names, and sheet names.
Because Excel is case insensitive (with the exception of string constants) we convert all input tokens to lowercase to map differently capitalized tokens to a single token, otherwise the same function (like \texttt{SUM} and~\texttt{sum}) will map to different tokens.
The final vocabulary size is 16,000 tokens.
As a full example,  \texttt{=SUMIF(B1:B5, "Not available", A1:A5)} is tokenized as
\texttt{= sumif ( b 1 : b 5 , \verbvisiblespace{}  " not \verbvisiblespace{} available "} \texttt{, \verbvisiblespace{} a 1 : a 5 )}
with space tokens denoted by \texttt{\verbvisiblespace{}}.

\subsection{Pre-training}
\label{approach:training}

We use a combination of two existing (tail masking and noisy auto-encoding) and two adapted pre-training objectives (language-aware masked span prediction and user-inspired denoising).
An overview is shown in Figure~\ref{fig:pretraining-objectives}.

\begin{description}
\item[Tail masking (TM)]We perform tail masking \cite{bart} on the character level, where the model has to predict the trailing \{30\%, 40\%, $\cdots$, 70\%\} of characters given the prefix.

\item[Language-aware masked span prediction (laMSP)]
We apply masked span prediction but force each span to respect Excel lexer boundaries.
For example, when an Excel cell reference \texttt{BC18} is divided into four tokens \texttt{B C 1 8}, we ensure that either all or none of its constituent tokens is masked.
Consecutive masked tokens are represented with a single \texttt{<mask>} token.
Inspired by Mixture-of-Denoisers \cite{ul2}, we mask spans of tokens using combinations of high (35\%) and low (15\%) masking rates, and long (6 tokens) and short (2 tokens) average span lengths.

\item[Random noise (RN)]
We randomly corrupt (insert, delete, or update) 10\% of tokens in the input sequence.

\item[User-inspired noise (UN)]
We introduce 17 noise operators that mirror mistakes that real users might make when writing Excel formulas. These operators are based on a combination of questions on help forums, our analysis of a user-study and our domain expertise.
For example, users often write formulas with the incorrect function arity for built-in functions such as \texttt{SUMIF}.
While pre-training, we randomly choose one of these operators to introduce noise into the input.
\end{description}
Following \citet{ul2} and \citet{chang2020pre}, rather than apply all pre-training objectives on every batch and then combine losses, we pick a single objective for each batch.
We use a higher probability for laMSP, as it contains four more combinations of high and low mask rates and long and short span lengths.
Empirically, we found these probabilities for objective sampling to perform well: 50\% laMSP, 20\% TM, 20\% UN, and 5\% RN.
We leave the sequence intact with a 5\% probability to mitigate the risk that the model learns to change all sequences (ignoring correctness/completeness).
\footnote{A technical appendix
with operator details and
evaluation dataset information is available at \url{https://github.com/microsoft/prose-benchmarks/tree/main/FLAME}}
\section{Downstream Tasks}
\label{subsection:downstreamtasks}
We consider three different downstream tasks: last-mile repair, formula completion and formula retrieval.

\subsection{Last-mile Repair}
Last-mile repair refers to repairs that require few edits and fix syntax and simple semantic errors, such as wrong function call arity~\cite{lamirage}. 
In this setting, \system{} is given the buggy formula as the input sequence, and the task is to generate the intended (valid) formula.

\begin{example} A wrong arity mistake in \texttt{ISERROR}.\\
\textbf{Buggy:} \texttt{=IF(ISERROR(G6 *1.2, ""\colorbox{myred}{)})}\\
\textbf{Repaired:} \texttt{=IF(ISERROR(G6 *1.2\colorbox{mygreen}{)}, "")}
\end{example} 

\paragraph{Fine-tuning}
We create a fine-tuning dataset for all systems by taking 200K well-formed formulas from Excel help forums\footnote{We use \url{mrexcel.com} and \url{excelforum.com} for curating fine-tuning dataset for last-mile repair task.} and randomly applying our user-inspired noise operators to generate broken formulas.

\paragraph{Evaluation}
We evaluate all systems on two benchmarks.
We use the 273 labeled Excel formulas used in recent last-mile repair literature~\cite{ring},
which was sourced from Excel help forums, and refer to this benchmark set as \textbf{\aaai{}}.
In addition, we reserve a split of randomly sampled 500 formulas derived using the same procedure as our fine-tuning dataset to create a \textbf{Synthetic} benchmark set.
We report the fraction of exact matches in top-1 and top-5 candidates for each system.
Before comparing, we normalize formulas by capitalizing cell references and identifiers, and removing whitespace tokens.

\subsection{Formula Completion} 
Code completion (given a prefix) is a popular code task for language models. We perform this task on formulas.

\begin{example} Formula completion.\\\label{autocompletionexample}%
\textbf{Prefix:} \texttt{=B2<=EDATE(}\\
\textbf{Completion:} \texttt{=B2<=EDATE(TODAY(),-33)}
\end{example} 

\paragraph{Fine-tuning}
We curated a fine-tuning dataset for completion by tokenizing 189k formulas and sampling a prefix length of $\{0.2, \cdots, 0.7, 0.8\}$ fraction of tokens.

\paragraph{Evaluation}
We evaluate completion on a single benchmark, consisting of the 273 ground truth formulas from the last-mile repair \aaai{} benchmark.
We predict completions given a \{0.5, 0.75, 0.90\}\% prefix.
Some parts of formulas are hard to predict without more context \cite{guo2021learning}, such as cell references, sheet names, string literals, and numbers.
Therefore, in addition to \textbf{exact match}, we also consider \textbf{sketch match} for completion with respect to the ground truth.
For example, in Example~\ref{autocompletionexample}, the number \texttt{-33} is highly contextual, so in a sketch we match with its token type (\texttt{number}).

\subsection{Similar Formula Retrieval}
A key task in data management is to
perform efficient information retrieval based on a user query.
In this downstream task, we embed a formula to retrieve similar formulas.
We use mean pooling to convert token embeddings from \system{}'s encoder to a fixed-size embedding \cite{sentencebert}.

\begin{example}
\label{retrieval}
Given three formulas
\begin{lstlisting}[numbers=none]
VLOOKUP($A1, $A$1:$AY$132, 42, FALSE)
VLOOKUP(P6, 'Other'!$A$3:$C$6, 3, FALSE)
C1-VLOOKUP(A1, 'F'!$A$3:$D$16, 4, FALSE)
\end{lstlisting}
the edit similarity is higher between the first two formulas compared to the 
first and third.
The embedding similarity should reflect this.
\end{example} 

\paragraph{Fine-tuning}
We sample 10K formulas from the public Enron spreadsheet corpus \cite{enron} and mask constants. 
We then compute all pairwise similarities using the Levenshtein edit similarity over lexer tokens.
Instead of fine-tuning the whole encoder, we train two fully connected layers that transform the embeddings of a pair of formulas to make their cosine similarity match the edit similarities \cite{adapter}.

\paragraph{Evaluation}
We evaluate formula retrieval on a collection of 1000 formulas from the Enron corpus---also pre-processed as in the previous section. We compare the Pearson's correlation between the pairwise cosine similarity over embeddings and the (desired, but expensive) target token edit similarity.

\section{Evaluation}

We perform experiments to answer the following research questions:

\newcommand{\rqperformance}[0]{RQ1}
\newcommand{\rqpretrain}[0]{RQ2}

\begin{itemize}
    \item \textbf{\rqperformance{}}: How does \system{} perform on formula repair, completion, and similarity formula retrieval compared to substantially larger language models?
    \item \textbf{\rqpretrain{}}: How do design decisions (data curation, model size, objectives, and tokenizer) affect \system{}'s downstream performance?
\end{itemize}

\begin{table}[tb]
\small
\centering
\caption{Architecture and size comparisons for models}
\begin{tabular}{@{}llr@{}}
\toprule
System        & Architecture    & \# parameters \\ \midrule
Codex-Cushman & Decoder         & 12 billion           \\
Codex-Davinci & Decoder         & 175 billion          \\
CodeT5 (base) & Encoder-Decoder & 220 million          \\
\system{} (ours)  & Encoder-Decoder & 60 million           \\ \bottomrule
\end{tabular}
\label{table:systems}
\end{table}

\paragraph{Baselines and Configurations}
We compare \system{} to the (much larger) models summarized in Table~\ref{table:systems}.
For Codex baselines, we use nucleus sampling \cite{holtzman2019curious} (temperature = 0.7) and sample 50 sequences per task.
We sort these sequences based on their average token log probabilities~\cite{ring}.
For CodeT5, we use beam search (width = 50), and we consider the top 50 sequences.

\subsection{Training Details}
\label{subsection:trainingdetails}
We pre-train \system{} for 10 epochs and fine-tune CodeT5 and \system{} on a cluster with 16 AMD MI200s, 96 cores and 900 GB RAM. 
We use an AdaFactor optimizer with 1e-4 learning rate, clip factor of 1.0, and a linear learning rate schedule with 100 warm-up steps.
We fine-tune \system{} for 
2
epochs for repair and completion and fine-tune CodeT5 for 25 epochs with a patience of 5 epochs. We fine-tune \system{} and others for 10 epochs for the formula retrieval experiments.
For fine-tuning, we use a weight decay of 0.1.
We carry out all Codex experiments on a cluster with 8 V100s, 40 cores, and 672 GB RAM.
For Codex fine-tuning, we use LoRA~\cite{lora}.

\subsection{\rqperformance{}: Downstream Performance}

\label{subsection:rq1}
We now compare \system{} to other language models, including substantially larger ones, on three formula intelligence tasks.

\subsubsection{Last-mile repair}

We fine-tune \system{}, CodeT5, and Cushman and use few-shot prompts with three shots for Davinci.
Even though noisy auto-encoding closely resembles last-mile repair, we find that fine-tuning \system{} helps direct it towards a particular task.

\begin{table}[tb]
\small
\centering
\caption{
   Performance for last-mile repair. The top section reports fine-tuned and few-shot (FS) results, the bottom section reports zero-shot (ZS) and un-fine-tuned results.
   With fine-tuning \system{} outperforms larger models in the \aaai{} benchmark at top-5, and comes in second at top-1.
   Without fine-tuning or examples, \system{} outperforms all other models.
(\textbf{best} and \underline{second best})
}
\begin{tabular}{llrrr}
\toprule
\multicolumn{1}{c}{}       & \multicolumn{2}{c}{Forum} & \multicolumn{2}{c}{Synthetic} \\ \cmidrule(lr){2-3} \cmidrule(lr){4-5}
\multicolumn{1}{c}{}       & T@1              & T@5              & T@1              & T@5             \\ \midrule
Cushman     & \textbf{0.79}    & \underline{0.88} & \textbf{0.87}             & \textbf{0.93}            \\
Davinci (FS)               & \underline{0.76} & \textbf{0.89}    & 0.54             & 0.77            \\
CodeT5-base        & 0.70             & 0.84             & \underline{0.84}             & 0.90            \\
CodeT5-small        & 0.72             & 0.83             & 0.82             & 0.89            \\
\system{}           & \underline{0.76} & \textbf{0.89}    & 0.83             & \underline{0.91}            \\ \midrule
Cushman (ZS) & 0.55          & \underline{0.85}          & 0.41          & 0.63          \\
Davinci (ZS) & \underline{0.60}          & 0.82          & \underline{0.51}          & \underline{0.75}          \\
\system{}   & \textbf{0.71} & \textbf{0.88} & \textbf{0.74} & \textbf{0.85}  \\ \bottomrule
\end{tabular}
\label{table:finetuned}
\end{table}

We summarize the results in Table~\ref{table:finetuned}. 
Overall, the \aaai{} benchmark is more difficult, with all models scoring lower at top-1 compared to the Synthetic benchmark (with the exception of Davinci, which may have observed more forum content in its training).
Because \system{}'s pre-training incorporates noise operations that are inspired by real-user mistakes, we find that \system{} 
achieves similar performance to the much larger Davinci model at both top-1 and top-5 on the \aaai{} benchmark.
Cushman, which we fine-tune on similar noise operations, shows the most consistently strong performance.
We also find that a general-purpose code model, like CodeT5, despite being fine-tuned on spreadsheet formula errors substantially lags on real user mistakes.

The bottom section of Table~\ref{table:finetuned} shows performance without fine-tuning of \system{}, and without examples for Codex.
Because of the denoising objectives, \system{} performs better than Codex baselines.
Especially at top-1, fine-tuning helps \system{} perform the right task.

\begin{figure*}[h]
  \centering
  \includegraphics[width=\linewidth]{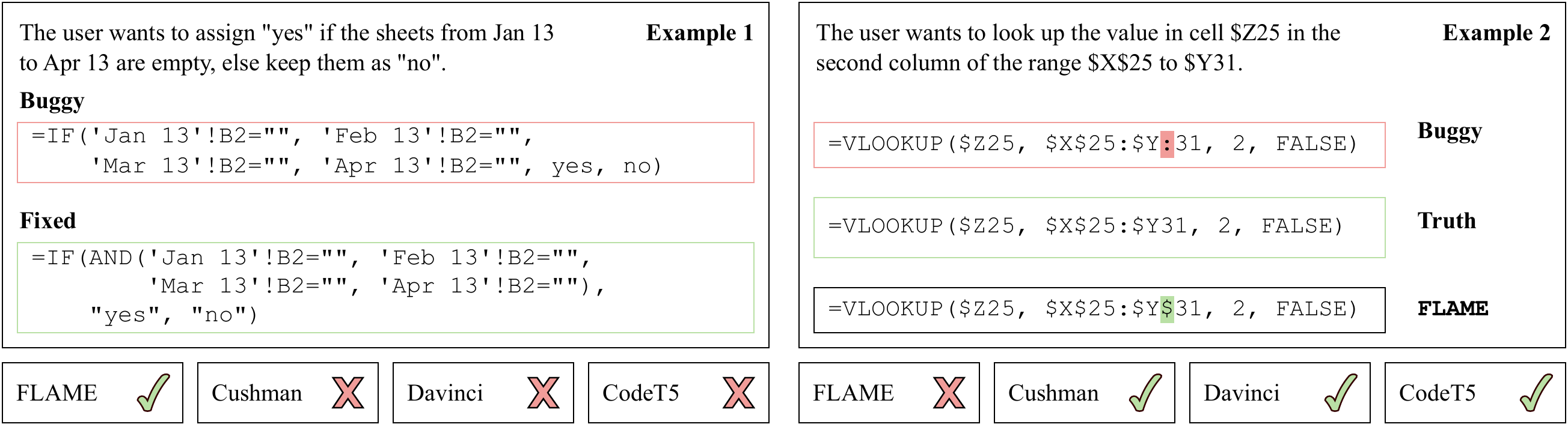}
  \caption{Repair tasks with diverging performance. In Example 1, the user did not use the \texttt{AND} function and missed double quotes around string literals \texttt{yes} and \texttt{no}. \system{} fixes this (in top-5), while other models fail. In Example 2, \system{}'s top candidate is syntactically valid but does not match the user's fix,
  while other models' predictions do.
  }
  \label{fig:example}
\end{figure*}

In Figure~\ref{fig:example}, we show examples where \system{} gets the correct fix and other models do not, and vice versa.
We note that in some cases, \system{}'s fixes appear to be more natural, but fail to match the user's ground truth repair.

\begin{figure}
    \centering
    \includegraphics[width=\linewidth, height=4.5cm]{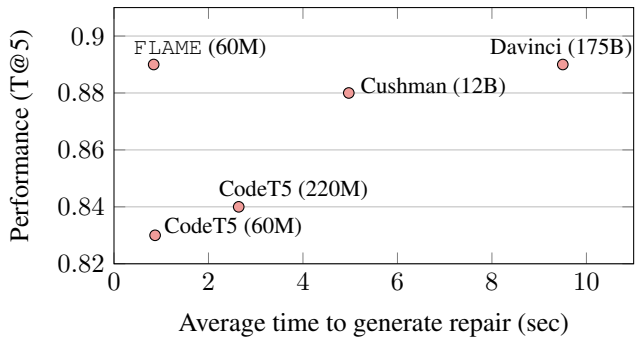}
    \caption{Last mile repair performance (top 5) compared to the average time (secs.) to generate repair candidates on the \aaai{} benchmark.}
    \label{fig:time_lmr}
\end{figure}

\system{}'s substantially smaller model size can facilitate model training and deployment.
To demonstrate this empirically, we consider the average
inference latency for the last-mile repair task. We focus
on this task as it can be directly translated to a user task (repairing formulas) and in contrast to completion, where model performance may vary based on the length of prefix, all models here have the same input. We carry out all measurements
on a CPU to emulate client-side deployment, where a GPU may
not be available. \Cref{fig:time_lmr} shows each model's last-mile repair rate (with a top-5 cutoff) and the average latency (seconds) to generate repair candidates for the Forum dataset. Cushman and Davinci results include network time as they are only available through an API. Davinci is used in a few-shot setting, and all other models are fine-tuned. CodeT5 (60M) and \system{} have the lowest, and comparable, latency. However, \system{} achieves a substantially higher repair rate compared to CodeT5.

\subsubsection{Formula completion}

The auto-regressive nature of Codex models and \system{}'s pre-training allows us to evaluate their zero-shot performance\footnote{We fine-tuned Codex-Cushman and \system{} but observed worse performance, possibly from over-fitting.} for formula completion.
Note that we fine-tune CodeT5 for this task as it is pre-trained on smaller span lengths (1 to 5 tokens) and generates special mask tokens (\texttt{<MASK1>}) in a zero-shot setting.
We compute exact match and sketch match with top-5 results.

\begin{table}[htb]
\small
\caption{Zero-shot completion for \system{}, Codex-Cushman and Codex-Davinci, and fine-tuned CodeT5.
Given $\{ 0.50, 0.75, 0.90 \}$ as formula prefix, we report the proportion of formulas completed in the top 5.
\system{} outperforms all the large language models in the exact match setting and most (2/3) of the sketch match settings. (\textbf{best} and \underline{second best}).}
\centering
\begin{tabular}{lrrrrrrrr}
\toprule
\multirow{2}{*}{\textbf{Model}} & \multicolumn{3}{c}{\textbf{Exact Match}} & \multicolumn{3}{c}{\textbf{Sketch Match}} \\ \cmidrule(l){2-4} \cmidrule(l){5-7} 
 & \textbf{0.50} & \textbf{0.75} & \textbf{0.90} & \textbf{0.50} & \textbf{0.75} & \textbf{0.90} \\ \midrule
Cushman & \underline{0.04} & 0.27 & 0.61 & \textbf{0.26} & 0.47 & 0.71 \\
Davinci & 0.03 & \underline{0.31} & \underline{0.64} & \underline{0.25} & \underline{0.53} & \underline{0.76} \\
CodeT5 & 0.02  & 0.10  & 0.27 & 0.09 & 0.20 & 0.39 \\
\system{} & \textbf{0.06} & \textbf{0.34} & \textbf{0.70} & 0.24 & \textbf{0.55} & \textbf{0.84} \\ \bottomrule
\end{tabular}
\label{table:autocompletion}
\end{table}

\begin{figure}[htb]
  \centering
\includegraphics[width=\columnwidth]{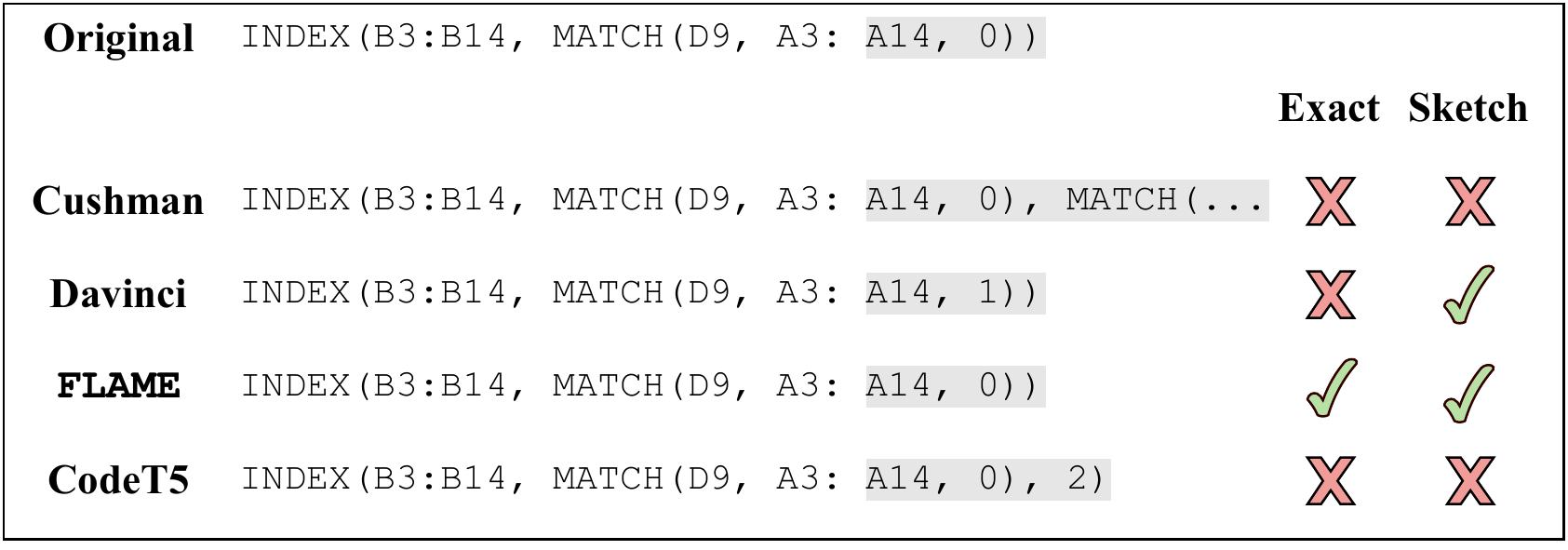}
  \caption{Completion for a 75\% prefix.
   All models correctly predict that the \texttt{MATCH} range goes from A3 to A14 based on the \texttt{INDEX} range B3:B14 in the prefix but result in (otherwise) different completions. \system{} generates the exact ground truth while Codex-Davinci generates a formula with same sketch as the ground truth. 
   }\label{fig:autocomplete_insight}
\end{figure}

Table~\ref{table:autocompletion} summarizes our results.
\system{} completions reflect correctly chosen cell references and other constants, and function names, and thus obtains a higher exact match rate across all prefix cutoffs.
When we consider sketch match, which ignores cell reference and constant differences, we find that larger models like Cushman and Davinci are competitive in a zero-shot setting.
For longer prefixes, they fall behind \system{}. %
CodeT5 performs poorly as it does not have either benefit: it is not a model with substantial number of parameters nor has it been designed for spreadsheet formulas.
Figure~\ref{fig:autocomplete_insight} shows an example where \system{} can complete with both exact and sketch match, while other systems fail on either or both criteria.

\begin{table*}[!t]
\small
\centering
\caption{
Impact of a smaller model size and global formula deduplication (versus our per-workbook deduplication) on top-1 Last-Mile Repair (LMR) and top-5 Autocompletion (AC) with Exact Match (EM) and Sketch Match (SM).
Smaller model performs worse across the board. Global deduplication reduces performance by up to 30 points.
}
\begin{tabular}{lrrrrrrrr}
\toprule
 & \multicolumn{6}{c}{\textbf{Zeroshot}} & \multicolumn{2}{c}{\textbf{Fine-tuned}} \\ \cmidrule(lr){2-7} \cmidrule(lr){8-9}
 & \multicolumn{2}{c}{\textbf{LMR}} & \multicolumn{2}{c}{\textbf{AC (EM)}} & \multicolumn{2}{c}{\textbf{AC (SM)}} & \multicolumn{2}{c}{\textbf{LMR}} \\ \cmidrule(lr){2-3} \cmidrule(lr){4-5} \cmidrule(lr){6-7} \cmidrule(lr){8-9}
 & \textbf{Forum} & \textbf{Synth} & \textbf{0.75} & \textbf{0.90} & \textbf{0.75} & \textbf{0.90} & \textbf{Forum} & \textbf{Synth} \\ \midrule
\system{} (60M) & \textbf{0.71} & \textbf{0.74} & \textbf{0.34} & \textbf{0.70} & \textbf{0.55} & \textbf{0.84} & \textbf{0.76} & \textbf{0.83} \\
\system{} (16M) & 0.68 & 0.64 & 0.24 & 0.59 & 0.54 & 0.76 & 0.73 & 0.78 \\
\system{} (60M w/ global deduplication) & 0.57 & 0.56 & 0.15 & 0.45 & 0.41 & 0.59 & 0.68 & 0.76 \\
\bottomrule
\end{tabular}
\label{table:rq3}
\end{table*}

\subsubsection{Formula retrieval}

For formula retrieval, we compare \system{} to
CodeT5, CodeBERT \cite{codebert}, and GraphCodeBERT (GCB) \cite{graphcodebert}.
Table~\ref{table:retrieval} shows the correlation between 
token-edit similarity and embeddings-based cosine
similarity.
\system{}'s embeddings result in a higher Pearson's correlation with our desired (but expensive) token-edit similarity metric, both without fine-tuning and with fine-tuning.
As a result of its small size, \system{} can also embed formulas up to 41\% faster than the next best model (CodeT5).
Surprisingly, CodeBERT and GraphCodeBERT, which were both trained on a code similarity objective, perform worse without fine-tuning.
We hypothesize that the encoder-decoder architecture,
shared by both CodeT5 and \system{}, leads to more effective formula embeddings.

\begin{table}[htb]
\small
\caption{
Correlation of cosine similarity over embeddings and 
token-edit similarity.
\system{}'s pre-trained embeddings are substantially more correlated with token-edit similarity.
After fine-tuning, both \system{} (60M) and CodeT5 (220M) perform best, but
\system{} embeds formulas up to 41\% faster.
}
\label{table:retrieval}
\begin{tabular}{lrrr}
\toprule
\textbf{Model} & \textbf{Pre-trained}   & \textbf{Fine-tuned} & \textbf{Time (ms)} \\ \midrule
\system{}                                       & \textbf{0.707}  & \textbf{0.904}           & \textbf{13.2}                                                                 \\
CodeT5                                      & 0.688  & 0.899           &  22.7                                                                 \\
CodeBERT                                    & 0.493  & 0.868           &  24.7                                                                 \\
GCB                               & 0.551  &  0.881           &  26.4                                                                 \\ \bottomrule
\end{tabular}
\end{table}

\subsection{\rqpretrain{}: Pre-training design decisions} \label{subsection:rq2}
We investigate  the impact of \system{}'s data curation,
model size, the use of domain-specific pre-training objectives, and domain-specific tokenizer.

\subsubsection{Training data curation} \label{sssec:trainingdata}
Previous work \cite{lee2021deduplicating,kandpal2022deduplicating} has shown that deduplication can improve 
language model performance.
We curated our pre-training dataset by 
performing sketch-based formula deduplication within each workbook.
Alternatively, one can perform global deduplication across all workbooks.
This results in a pre-training set of 591K formulas.
Table~\ref{table:rq3} shows that training on this smaller corpus results in a worse model (e.g., -8pp on fine-tuned repair, -19pp on exact match completion).
While global deduplication allows for faster training, the resulting model observes less naturally occurring repetition across users' spreadsheets.

\subsubsection{Model size}
Table~\ref{table:rq3} compares \system{} with 16M parameters to the original 60M parameters.
We find that performance declines slightly across benchmarks when we reduce model size. However, we note that \system{}-16M still outperforms both Codex models in some tasks, such as zero-shot last-mile repair at top-1 on \aaai{} benchmark, highlighting the potential of leveraging smaller models in specific domains. Note that \system{}-16M is pretrained using the same approach as \system{}-60M with the same amount of data and then finetuned for respective downstream tasks.

\begin{table}[htb]
\small
\caption{We incrementally add laMSP (M), user-inspired denoising (N) and Excel-specific tokenizer (T) to T5. The final row represents \system{}. We evaluate on last-mile repair (LMR) and completion (C) with 75\% and 90\% prefixes for Forum benchmark and report exact (EM) and sketch match (SM). We observe improvements across settings. * is \system{}.
}
\label{table:ablation}
\centering
\begin{tabular}{@{}l@{}rrrrrr@{}}
\toprule
\textbf{Ablations} & \multicolumn{2}{c}{\textbf{LMR}}  & \multicolumn{2}{c}{\textbf{C (EM)}} & \multicolumn{2}{c}{\textbf{C (SM)}}  \\ \cmidrule(lr){2-3} \cmidrule(lr){4-5} \cmidrule(lr){6-7}
                                          & \textbf{T@1}  & \textbf{T@5}  & \textbf{0.75}    & \textbf{0.90} & \textbf{0.75}    & \textbf{0.90} \\ \midrule
T5            & 0.67          & 0.81          & 0.07             & 0.22          & 0.25             & 0.37          \\
+ M & 0.73          & 0.85          & 0.15             & 0.51          & 0.35             & 0.54          \\
+ M + N & 0.75          & 0.87          & 0.16             & 0.51          & 0.34             & 0.59          \\
+ M + N + T* & \textbf{0.76} & \textbf{0.89} & \textbf{0.34}    & \textbf{0.70} & \textbf{0.55}    & \textbf{0.84} \\
\bottomrule
\end{tabular}
\end{table}

\subsubsection{Pre-training objectives and tokenizer}

To evaluate the impact of each  pre-training objective and domain-specific tokenization, we incrementally add these to a T5 model.
Table~\ref{table:ablation} summarizes these results.

Language-aware MSP (laMSP) causes the most notable improvements (up to 29 percentage points) across all downstream tasks.  laMSP forces the noisy auto-encoding task to recover full Excel tokens without leveraging partial tokens, which can leak information, for example, in cell ranges.
We then add user inspired denoising, which provides additional
benefits for last-mile repair (which is a more difficult 
version of noisy auto-encoding).
Finally, the Excel-specific tokenizer yields improvements across the board, but it specifically shines for completion.
We attribute this improvement to the consistent tokenization of strings with different capitalization using the Excel-specific tokenizer. 
For example, we found base T5 to struggle with spreadsheet names and built-in function names, generating code such as \texttt{Sum!C3} instead of \texttt{SUM(C3)}, when given a non-upper-cased \texttt{sum}.

\paragraph{Future Work} We provide three pointers for future work.
First, fast inference of our smaller model provides an
opportunity for more complex search strategies that
incorporate symbolic constraints during decoding~\cite{synchromesh}.
Second, \system{} embeddings, which we showed are effective
for formula similarity, should be explored for embedding tables with formula content~\cite{spreadsheetcoder,tapex}.
Third, further innovation in tokenization (e.g. 
separately encoding formula tokens and constants) may help
drive model sizes even lower.

\section{Conclusions}
We present \system{}, a small (60M parameter) language model
for spreadsheet formulas, designed with domain-specific
data curation, tokenization, and pre-training objectives.
We implemented \system{} for Excel formulas and evaluate
on last-mile repair, auto-completion, and similar formula retrieval. 
We compare to the much larger CodeT5, Codex-Cushman, and Codex-Davinci for repair and completion,
and to CodeT5, CodeBERT, and GraphCodeBERT for
similar formula retrieval.
\system{} obtains the best results in 11 out of
16 evaluation settings, competitively
fixing, completing, and retrieving formulas.

\section*{Acknowledgments}

We thank Microsoft Research Cambridge for sharing insights and data.
We thank OCTO at Microsoft (in particular Gopi Kumar and the AMD vTeam) for providing us with compute resources.
We also thank the Excel team for their feedback and encouragement in pursuing this work.

\bibliography{ref.bib}

\clearpage
\appendix
\section{Technical Appendix}
\section{Noise operators}\label{appendix:noise}
We detail the 17 noise operators used during
\system{}'s pretraining.

\begin{enumerate}
    \item \textbf{Wrong range:} we replace the range operator colon (\texttt{:}), with one of the following symbols: $\{\texttt{; , space "}\}$, or we delete the range operator.
    \item \textbf{Malformed range:} A range consists of 4 elements: \texttt{col1}, \texttt{row1}, \texttt{col2}, \texttt{row2} written as \texttt{col1row1:col2row2}. We randomly delete one of these elements, e.g. \texttt{col1:col2row2}
    \item \textbf{Space between function and arguments in a call}: We introduce a space between the function name and the opening parentheses for built-in functions. For example: we transform \texttt{SUM(A1:A10)} to \texttt{SUM (A1:A10)}
    \item \textbf{Change number of arguments:} We change the number of arguments for functions with fixed function arity. For example, \texttt{IF} has a minimum arity of 2 and maximum arity of 3. Specifically, if a function contains arguments equal to its minimum function arity, then we randomly delete one argument. Whereas if the number of arguments in a call match the function's max arity,
    we randomly copy one of the existing arguments and append as a new argument in the call to break the arity limit.
    For example, \texttt{IF(A2>10, True, False)} can become \texttt{IF(A2>10, True, False, False)}
    \item \textbf{Swap arguments:} If a function takes different types of arguments, then we swap these arguments. For example: \texttt{IF(A1>10, 1, 2)} can become \texttt{IF(1, A1>10, 2)}.
    \item \textbf{Space between relational operators: } We add space between relational operators, such as \texttt{< =}.
    \item \textbf{Swap relational operators: } We swap relational operators. For example, \texttt{<=} becomes \texttt{=<}.
    \item \textbf{Inequality noise operator: } In Excel \texttt{<>} is the inequality operator. We replace this with the incorrect \texttt{!=} or \texttt{=!}.
    \item \textbf{Invalid equality:} We corrupt the equality operator. The equality operator in Excel is \texttt{=}, we replace it with \texttt{==} or \texttt{===}.
    \item \textbf{Malformed sheet name:} Multi-word sheet names in Excel need to be enclosed within single quotes (\texttt{'<sheet name>'}). We randomly choose to either delete the single quotes or replace them with double quotes. For example, \text{'Sheet 1'!A10} can become \texttt{"Sheet 1"!A10}.
    \item \textbf{Remove exclamation mark: } In Excel, sheet names are followed by an exclamation mark to denote sheet reference. We delete this exclamation mark.
    \item \textbf{Malformed strings: }We corrupt strings by either deleting the double quotes or replacing them with single quotes.
    \item \textbf{Add comma and remove parentheses:} We randomly choose to either insert a comma before a closing parenthesis or insert a comma and delete the closing parentheses.
    \item \textbf{Add random operators:} We define a set of operators that we randomly insert into the formula at a random position. These operators are:
    $\{$\texttt{+ - * / \^\ \& < > = . ) \#}$\}$ 
    \item \textbf{Add operator at the end: } We randomly add one of the operators mentioned previously at the end of the sequence.
    \item \textbf{Add parentheses}: We add opening and closing parentheses at random places.
    \item \textbf{Corrupting unreliable tokens: }Following \citet{lamirage}, we randomly add, delete or replace unreliable tokens. 
    Unreliable tokens are tokens where users often make mistakes, defined to be delimiters.   
\end{enumerate}

\section{Codex Prompts}\label{appendix:prompts}

For all our Codex experiments, we use the following prompts for zeroshot and finetuning and use a temperature of 0.7 based on ~\citet{ring}.

\noindent\textbf{Last Mile Repair}
    \begin{verbatim}
    ##### Fix bugs in the below code
    ### Buggy Excel
    <Buggy Formula>
    ### Fixed Excel
    <Fixed Ground Truth Formula>\end{verbatim}
\noindent\textbf{Autocompletion}  
    \begin{verbatim}
    ### Excel Formula
    <Partial Excel Formula>
    <Formula>
    ### Excel Formula
    IF(FALSE,NA(\end{verbatim}

\begin{figure}[!h]
    \centering
    \includegraphics[scale=0.5]{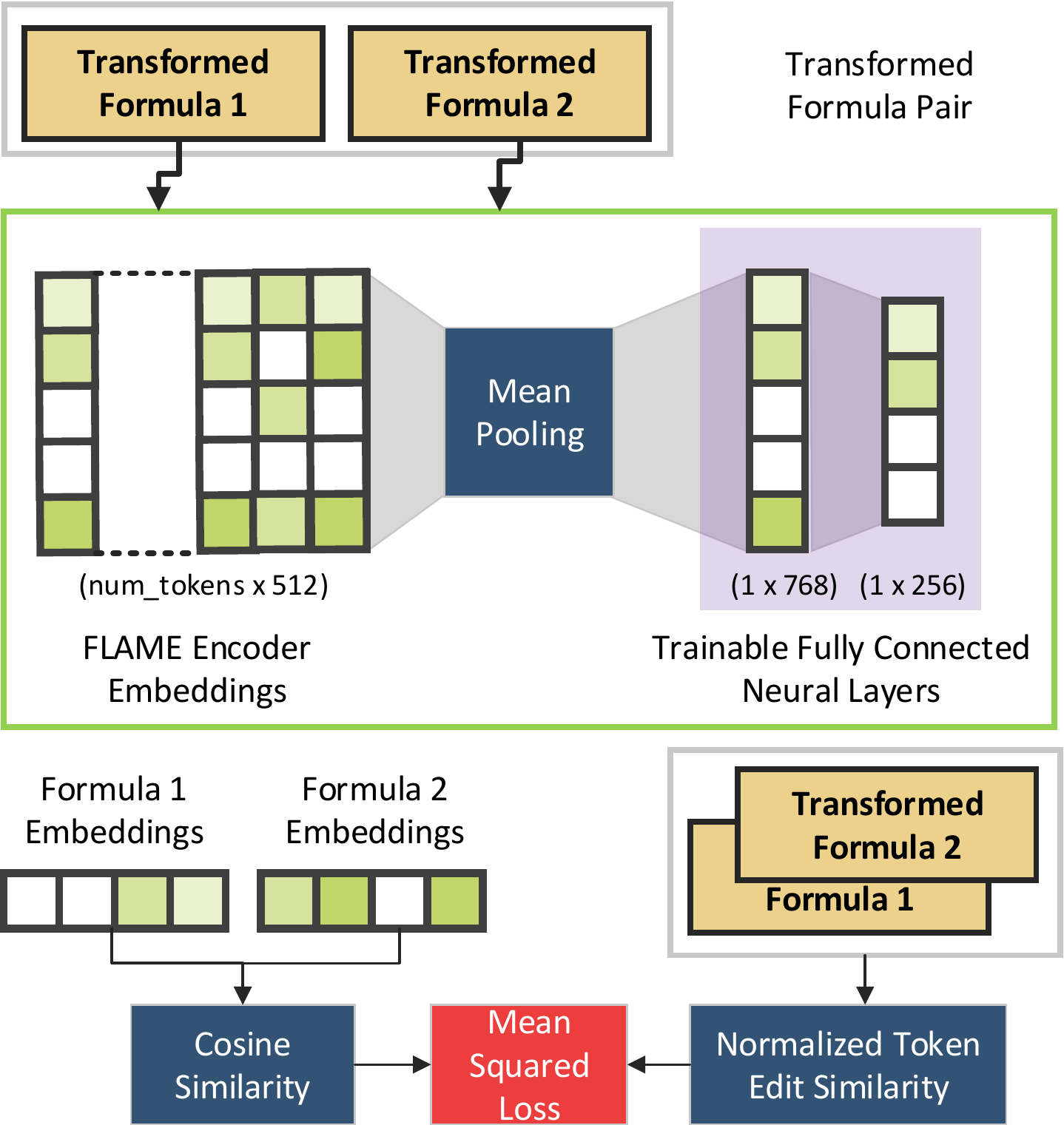}
    \caption{To better align formula embeddings to token edit similarity, we train two fully connected layers on top of the mean-pooled formula embedding. Finally, we compute the cosine similarity between two formulas and minimize mean-squared error to the token edit similarity.
    }
    \label{fig:embedding-finetuning}
\end{figure}

\begin{table*}[!htbp]
\caption{
Performance by decoder strategy for Autocompletion (top 5) with Exact Match and Sketch Match. For shorter prefixes, TopK Sampling performs better than other baselines in the Sketch Match metric. Whereas for longer prefixes Group Beam Search performs best in Sketch Match and outperforms others in Exact Match metric for all prefix lengths.}
\centering
\begin{tabular}{@{}lrrrrrrrrrr@{}}
\toprule
\multicolumn{1}{c}{\multirow{2}{*}{\textbf{Models}}} & \multicolumn{5}{c}{\textbf{Exact Match}}                                                         & \multicolumn{5}{c}{\textbf{Sketch Match}}                                                         \\ \cmidrule(l){2-11} 
\multicolumn{1}{c}{}                                 & \textbf{0.25} & \textbf{0.50} & \textbf{0.75} & \textbf{0.90} & \multicolumn{1}{r}{\textbf{0.99}} & \textbf{0.25} & \textbf{0.50} & \textbf{0.75} & \textbf{0.90} & \multicolumn{1}{r}{\textbf{0.99}} \\ \midrule
Beam Search                                                 & 0.00             & 0.06         & 0.33          & \textbf{0.71}         & 0.92                               & 0.10           & 0.25         & 0.54          & 0.82         & 0.94                               \\
Group Beam Search (groups $=2$)                                          & \textbf{0.01}          & \textbf{0.06}         & \textbf{0.34}          & 0.70          & \textbf{0.93}           & 0.10           & 0.24         & 0.55          & \textbf{0.84}         & \textbf{0.94}           \\
Nucleus Sampling                                             & 0.00             & 0.04         & 0.26          & 0.59         & 0.92                               & 0.14          & 0.30          & 0.53          & 0.74         & 0.92                               \\
TopK Sampling                                                & 0.00             & 0.04         & 0.25          & 0.62         & 0.92                               & \textbf{0.15}          & \textbf{0.30}         & \textbf{0.55}          & 0.76         & 0.92                               \\ \bottomrule
\end{tabular}
\label{appendix:decoderautocomplete}
\end{table*}

\section{Formula Retrieval Setup}
To improve the correlation of 
formula-embedding-based cosine similarity and
token-edit similarity (which is our target, but expensive), we add two fully connected layers on top of the
formula embedding, with a ReLU activation in between, as shown in 
Figure~\ref{fig:embedding-finetuning}.
We only finetune these two layers. We finetune by minimizing the
mean squared error between the cosine similarity over output vectors and the token-edit similarity.

To fine-tune, we sample 10000 formula pairs from the Enron dataset and train these two layers for 10 epochs using Adam and a learning rate of $1e^{-5}$. For the test dataset, we make sure that none of the formulas were seen while training the layers.
We note that finetuning these layers for \system{} took only 5 minutes and 43 secs on a single core i7-10610U CPU @ 1.8Ghz.

\section{Decoder Experiments}

We found that different decoding strategies work well with \system{} for different downstream tasks. 
Table~\ref{appendix:decoderautocomplete}, details the autocompletion results for different decoding strategies. 
For auto-completion with sketch match, top-k sampling shows superior performance up to 75\% of the prefix length.
We believe this is because autocompletion requires more diverse results that can be sampled using Nucleus and Top-K sampling techniques, particularly at shorter prefixes. 
However, Group Beam Search outperforms all the other techniques for the Exact Match metric and for longer prefixes ($>75$\%) in Sketch Match.

In Table~\ref{table:decoder}, we evaluate \system{}'s
last-mile repair performance using four different decoding strategies, Beam Search, Group Beam Search~\cite{vijayakumar2016diverse}, Nucleus Sampling~\cite{holtzman2019curious} and Top K Sampling~\cite{fan2018hierarchical}. 
We find \system{} to perform best with group beam search decoding, with a group size of 2.

\begin{table}[!htbp]
\caption{Performance by decoder strategy for last mile repair (LMR)
}

\centering
\begin{tblr}{
  column{2} = {r},
  column{3} = {r},
  hline{1,6} = {-}{0.08em},
  hline{2} = {-}{0.08em},
}
\textbf{Decoding Method} & \textbf{T@1}  & \textbf{T@5}  \\
Beam Search              & 0.76          & 0.88          \\
Group Beam Search        & \textbf{0.76} & \textbf{0.89} \\
Nucleus Sampling         & 0.72          & 0.85          \\
Top K Sampling           & 0.67          & 0.86          
\end{tblr}
\label{table:decoder}
\end{table}

\end{document}